\documentclass[sigconf]{acmart}

\usepackage{caption}
\usepackage{subcaption}
\usepackage{color}
\definecolor{orange}{RGB}{255,127,0}
\definecolor{limegreen}{RGB}{50, 205, 50}
\definecolor{violet}{RGB}{148,0,211}

\newif\ifCOMMENTS
\COMMENTStrue
\ifCOMMENTS

\else

\fi

\AtBeginDocument{%
  \providecommand\BibTeX{{%
    \normalfont B\kern-0.5em{\scshape i\kern-0.25em b}\kern-0.8em\TeX}}}

\setcopyright{acmcopyright}
\copyrightyear{2018}
\acmYear{2018}
\acmDOI{10.1145/1122445.1122456}

\acmConference[Woodstock '18]{Woodstock '18: ACM Symposium on Neural
  Gaze Detection}{June 03--05, 2018}{Woodstock, NY}
\acmBooktitle{Woodstock '18: ACM Symposium on Neural Gaze Detection,
  June 03--05, 2018, Woodstock, NY}
\acmPrice{15.00}
\acmISBN{978-1-4503-XXXX-X/18/06}

\copyrightyear{2021}
\acmYear{2021}
\setcopyright{rightsretained}
\acmConference[UIST '21 Adjunct]{The Adjunct Publication of the 34th Annual ACM Symposium on User Interface Software and Technology}{October 10--14, 2021}{Virtual Event, USA}
\acmBooktitle{The Adjunct Publication of the 34th Annual ACM Symposium on User Interface Software and Technology (UIST '21 Adjunct), October 10--14, 2021, Virtual Event, USA}\acmDOI{10.1145/3474349.3480194}
\acmISBN{978-1-4503-8655-5/21/10}

\begin{document}

\title[Enhancing Model Assessment in Vision-based IMT through Real-time Saliency Map Visualization]{Enhancing Model Assessment in Vision-based Interactive Machine Teaching through Real-time Saliency Map Visualization}



\author{Zhongyi Zhou, Koji Yatani}
\affiliation{
  \institution{Interactive Intelligent Systems Lab., The University of Tokyo}
  \city{Tokyo}
  \country{Japan}
}
\email{{zhongyi, koji}@iis-lab.org}


\begin{abstract}
Interactive Machine Teaching systems allow users to create customized machine learning models through an iterative process of user-guided training and model assessment.
They primarily offer confidence scores of each label or class as feedback for assessment by users.
However, we observe that such feedback does not necessarily suffice for users to confirm the behavior of the model.
In particular, confidence scores do not always offer the full understanding of what features in the data are used for learning, potentially leading to the creation of an incorrectly-trained model.
In this demonstration paper, we present a vision-based interactive machine teaching interface with real-time saliency map visualization in the assessment phase.
This visualization can offer feedback on which regions of each image frame the current model utilizes for classification, thus better guiding users to correct the corresponding concepts in the iterative teaching.

\end{abstract}




\maketitle

\section{Introduction}
Although Machine Learning (ML) has solved considerable research challenges, non-ML-experts can barely utilize this technology to benefit their life due to the lack of knowledge.
Recent studies in Interactive Machine Teaching (IMT)~\cite{ramos2020imt} have increased the accessibility of ML by creating various interactive tools for users to easily build an ML model.
For example, Teachable Machine~\cite{Carney2020teachable} is a web-based tool that allows users to teach a vision-based ML model by simply providing tens of examples in front of the web camera.
Similar to other tools like lobe.ai~\cite{lobeai}, existing interfaces primarily show a prediction confidence score of each label or class when the user is evaluating the trained model.
We argue that such feedback can be insufficient for users to correctly evaluate the model, which may cause unexpected results in its actual use.

In this paper, we present enhancement of a vision-based IMT system\footnote{This source code of this project is available at \color{blue}{\url{https://github.com/IIS-Lab/imt_vis_uist21}}} that can offer feedback using saliency maps during assessment by users.
Unlike the approach in Alqaraawi et al.'s study~\cite{Alqaraawi2020Evaluating} that visualizes the low-level features like edges, our interface superimposes the results of the importance scores of regions in a frame, highlighting which portion of the frame the current model is weighing for classification.
This visualization can help users confirm the behavior of the current model and identify potential issues.
For example, even though the model shows a high confidence score, it may utilize features taken from irrelevant regions in a video frame (e.g., objects and colors in a background).
Our interface allows users to find out such issues through the saliency maps.
We thus believe that it can make the train-feedback-correct cycle of interactive machine teaching more informed, contributing to more rapid development of a model.
\begin{figure*}[t]
    \centering
    \begin{subfigure}[b]{0.33\linewidth}
        \includegraphics[width=\textwidth]{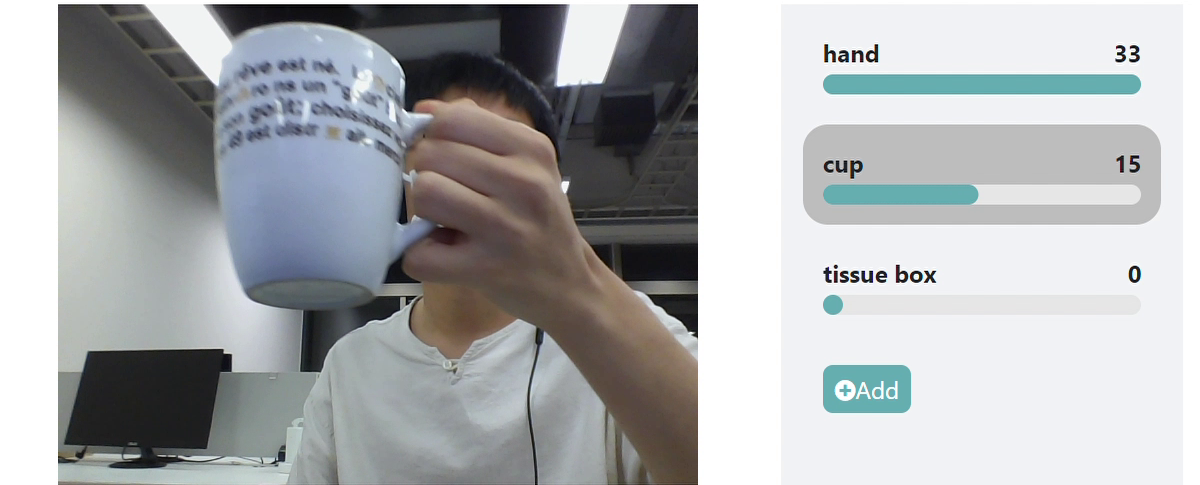}
        \caption{Teaching stage.}
        \label{fig: interface_input}
    \end{subfigure}
    \begin{subfigure}[b]{0.33\linewidth}
        \includegraphics[width=\textwidth]{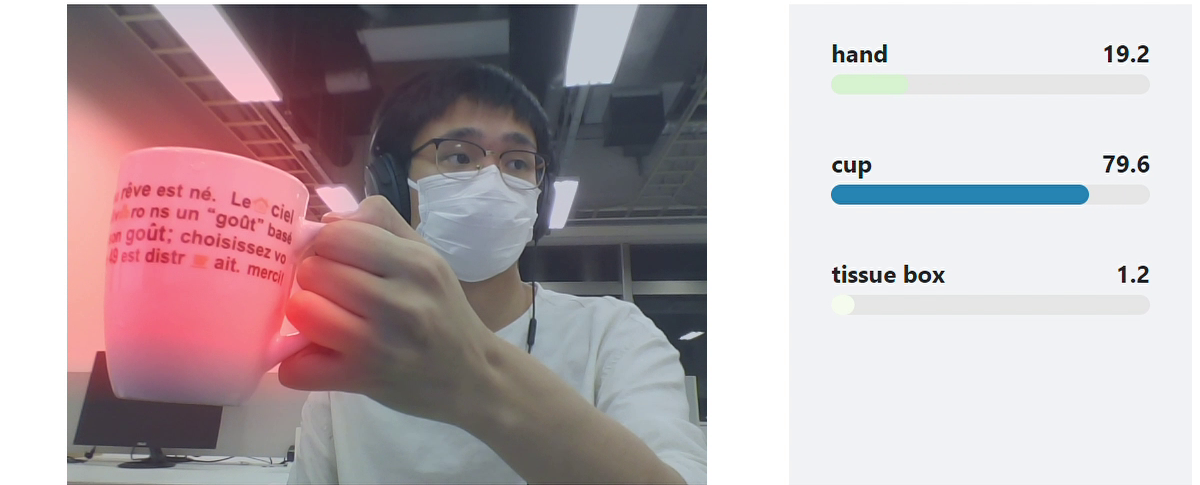}
        \caption{Evaluation stage.}
        \label{fig: interface_vis}
    \end{subfigure}
    \begin{subfigure}[b]{0.33\linewidth}
        \includegraphics[width=\textwidth]{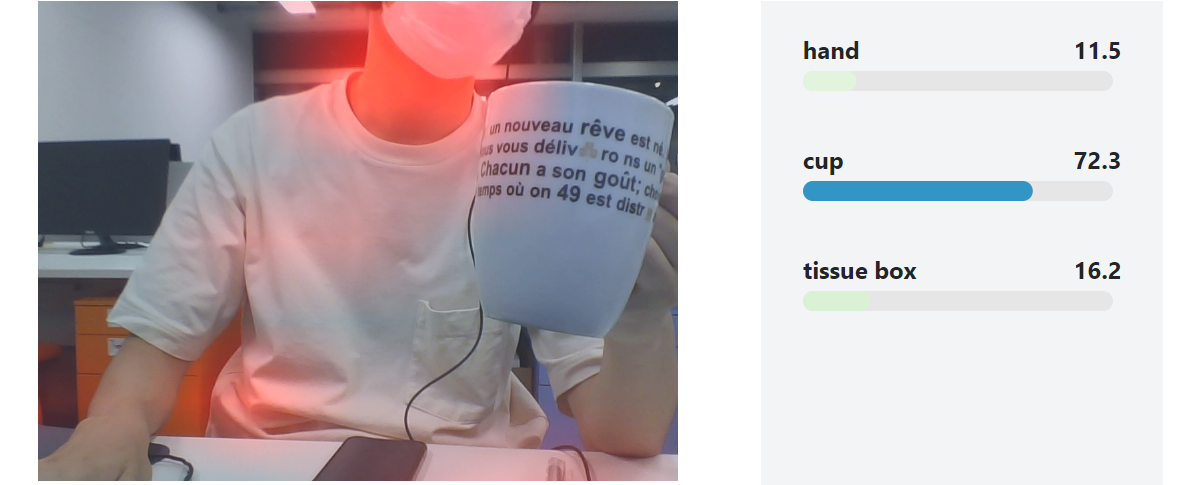}
        \caption{Wrong reference.}
        \label{fig: interface_wrong}
    \end{subfigure}
    \caption{System Interfaces. In \ref{fig: interface_input}, the user is teaching an ML model with samples of ``cup''. The interface counts and visualizes the number of teaching samples on the right side. \ref{fig: interface_vis} presents the interface when the person is evaluating the model in real time. The saliency map overlay is shown on the left side, and confidence scores of all the labels are on the right side. \ref{fig: interface_wrong} shows a case when the model is referring to a wrong object for a high-confidence correct prediction.}
    \label{fig: interface}
\end{figure*}

\begin{figure}[t]
    \centering
    \begin{subfigure}[b]{0.22\linewidth}
        \includegraphics[width=\textwidth]{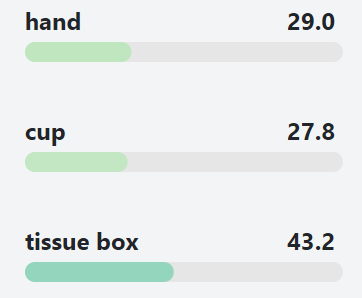}
        \caption{Scores}
        \label{fig: confusion_confidence}
    \end{subfigure}
    \begin{subfigure}[b]{0.25\linewidth}
        \includegraphics[width=\textwidth]{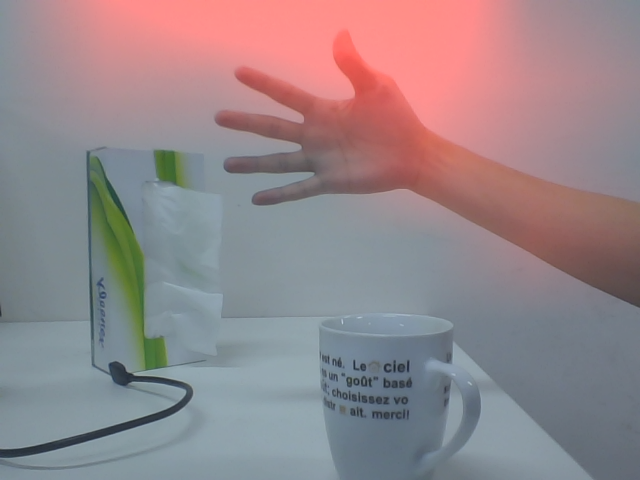}
        \caption{Hand}
        \label{fig: confusion_hand}
    \end{subfigure}
    \begin{subfigure}[b]{0.25\linewidth}
        \includegraphics[width=\textwidth]{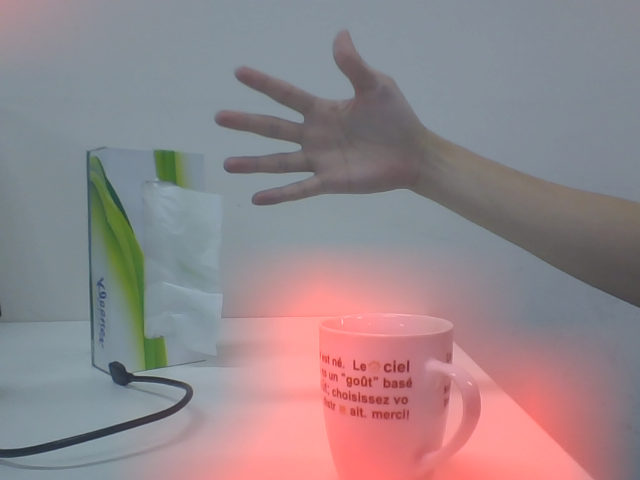}
        \caption{Cup}
        \label{fig: confusion_cup}
    \end{subfigure}
    \begin{subfigure}[b]{0.25\linewidth}
        \includegraphics[width=\textwidth]{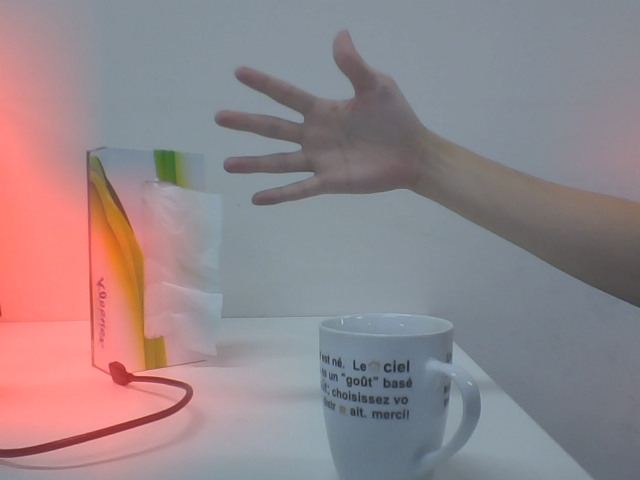}
        \caption{Tissue box}
        \label{fig: confusion_tissue}
    \end{subfigure}
    \caption{Saliency maps of different labels for the same image. Confidence scores blocks in \ref{fig: confusion_confidence} are clickable, and once clicked, it will activate the saliency map of the corresponding class (\ref{fig: confusion_hand}, \ref{fig: confusion_cup}, or \ref{fig: confusion_tissue}). If no class is specified, the system shows the saliency map of the class with the highest confidence score by default (\ref{fig: confusion_tissue}). }
    \label{fig: confusion}
\end{figure}


\section{Real-time Saliency Map during Model Assessment}

\subsection{Interface Design}
Our interface resembles existing vision-based IML systems.
Users can perform training by demonstrations in front of a camera. 
Figure~\ref{fig: interface_input} shows an example of the interface when a user is providing samples of a ``cup'' class.
After the user finishes providing the samples for all labels, the system trains a neural network using these data.
When the model is ready, the user can evaluate the model by performing another demonstration in front of the camera.
Figure~\ref{fig: interface_vis} shows an exemplary interface during the model assessment. 
The system provides two kinds of feedback in real time: a confidence score for each label and a saliency-map overlay on the video stream.
In Figure~\ref{fig: interface_vis}, the model shows the confidence score of $79.6\%$ for the cup class.
The saliency map provides the regions of the video frame that contributes to the classification of that particular class.
In this example, the model is correctly looking at a cup to perform classification.
Note that the system shows the saliency map of the class with the highest confidence score by default.
Users can also specify one class of interest for visualization.
For example, Figure~\ref{fig: confusion_hand}, \ref{fig: confusion_cup}, and \ref{fig: confusion_tissue} show the saliency map of each class if the user click the corresponding class in Figure~\ref{fig: confusion_confidence}.

\subsection{Implementation}

The system utilizes transfer learning to quickly train the network with a limited size of data.
Specifically, the system resizes each video frame into $224 \times 224$, and feeds the image into MobileNet$\_$v1~\cite{mobilenets2017andrew} pre-trained on ImageNet~\cite{russakovsky2015imagenet}.
This constitutes a feature vector for each image, and the system only uses it instead of the original images to reduce the training time. 
We attach a fully-connected layer to the output feature of MobileNet 
that transforms the features into logits, followed by a soft-max layer for classification.
This implementation also enables the saliency map shown above (the details of its implementation are explained in the next paragraph).
After the user finishes teaching, the system trains the one-layer network for ten epochs using the cross-entropy loss function with Adam Optimizer~\cite{adam2015Diederik} with default parameters in TensorFlow.js.

The system then shows saliency maps after the network finishes training.
We use Zhou et al.'s method~\cite{cam2016zhou} on the front end to compute the semantic-level saliency map of each video frame in real time.
Intuitively, the method computes importance scores at different image regions for making each prediction.
One critical benefit of the method is that it helps users easily understand which part of the image is contributing to a particular prediction made by the current model.
Our system leverages this property for users to explore different saliency maps through simple clicks (Figure~\ref{fig: confusion}).

\section{Preliminary Explorations and Results}
To understand the practical usability of our system, we tested the speed performance of our system on two device conditions, i.e., Integrated Graphics (IG) or Dedicated Graphics (DG).
We chose Intel(R) HD 630 as our IG card and NVIDIA GTX 1050 (Notebook) as the DG card for the experiment, representing an average-level property of current laptops.
The web-camera resolution was $640 \times 480$.
Training 3 classes (30 samples per class) took 3342 msecs on IG and 1943 msecs on DG.
The total latency for saliency map visualization was 113 and 51 msecs on IG and DG, respectively.
This latency can be broken down to the inference (deriving both confidence scores and saliency map values) and visualization rendering.
They were 45 and 68 msecs in IG while 15 and 36 msecs in DG.
These results suggest that our system can run in real time in commodity laptops.

We also conducted an informal case study to understand its potential benefits.
We found that the saliency map usually provided reasonable explanations for the predictions, confirming that the model was trained as expected. 
In occasional cases, however, the model tended to be overconfident of a correct prediction with wrong references.
Figure~\ref{fig: interface_wrong} shows an example in which the model was highly confident ($72.3\%$) that the user was presenting a ``cup'' (which is correct) by weighing the neck and the mask of a person.
Because the user taught the model by not only showing the ``cup'' through the web camera but also including other noisy information (e.g., the mask and the neck in this example), it is reasonable for the ML model to misunderstand the noisy information as important features of the ``cup''.
Such issues can occur more frequently if users perform teaching with a background with complex textures.
Users may notice similar failure when they change the perspective and/or position of the camera (see the difference between Figure~\ref{fig: interface_vis} and~\ref{fig: interface_wrong}).
Informing users of such failure through visualization is crucial so that they can identify potential causes and fix the issues in a later teaching phase.
Our visualization can enhance users' assessment of a model and lead to a more efficient cycle of developing a model through Interactive Machine Teaching.

\section{Conclusion}
In this paper, we present enhancement for model assessment in a vision-based interactive machine teaching.
In our current prototype, we employ saliency map visualization on a video stream along with confidence scores of classification.
Our preliminary exploration discovers the benefits of such visualization.
We plan to further explore different designs of visualization and conduct formal user studies to understand how such interfaces can accelerate the entire interactive machine teaching process.

\bibliographystyle{ACM-Reference-Format}
\bibliography{contents/reference}

\end{document}